\documentclass[conference]{IEEEtran}
\IEEEoverridecommandlockouts
\usepackage{cite}
\usepackage{amsmath,amssymb,amsfonts}
\usepackage{graphicx}
\usepackage{textcomp}
\usepackage{xcolor}
\usepackage{cite}
\usepackage{balance}
\usepackage{bm}
\usepackage{subfigure}
\usepackage{algorithm} 
\usepackage{algorithmic}  
\usepackage[algo2e]{algorithm2e} 
\usepackage{savesym}
\usepackage{siunitx}
\def\BibTeX{{\rm B\kern-.05em{\sc i\kern-.025em b}\kern-.08em
    T\kern-.1667em\lower.7ex\hbox{E}\kern-.125emX}}

\newcommand{\CLASSINPUTtoptextmargin}{2.4cm}
\begin{document}

\bstctlcite{IEEEexample:BSTcontrol}

%\title{Spider RIS: Dynamic Movable Reconfigurable Intelligent Surfaces for Enhanced Indoor Communication Systems

\title{Angular-Based Hybrid Beamforming for Wideband THz Massive MIMO Systems: Mitigating Beam Split by Leveraging Angular Spread \\

%\thanks{The work of Tho Le-Ngoc was also supported in part by the Natural Sciences and Engineering Research Council of Canada (NSERC), InterDigital Canada, and Prompt Quebec under an NSERC Alliance Grant.}
}

\author{\IEEEauthorblockN{Ibrahim Yildirim, Tho Le-Ngoc}
	\IEEEauthorblockA{Department of Electrical and Computer Engineering, McGill University, Montreal, QC, Canada \\
		Email: ibrahim.yildirim@mcgill.ca,
		tho.le-ngoc@mcgill.ca
		\vspace{-3ex}}
}

\maketitle

\begin{abstract}
Beam split is a critical challenge in wideband THz massive MIMO systems, arising from frequency-dependent beam misalignment that degrades communication performance, particularly in scenarios with narrow beamwidths and large arrays. This work proposes an angular-based hybrid beamforming framework that leverages angular spread to mitigate the beam split effect. Instead of relying on precise angular spread modeling, we utilize coarse angular information to guide the design of subcarrier-specific beams, effectively reducing misalignment across subcarriers. By broadening the effective beamwidth through angular spread, the proposed method enhances user coverage and alleviates beam split without requiring complex time-delay units or hardware-intensive solutions. Simulation results demonstrate that the proposed approach achieves significant improvements in spectral efficiency and beamforming accuracy while maintaining low computational and hardware complexity. This work provides a practical and efficient solution for addressing beam split in next-generation wideband THz communication systems.
\end{abstract}

\vspace{-2ex}
\section{Introduction}
As wireless communication progresses toward 6G, Terahertz (THz) communication has emerged as a key enabler due to its potential for ultra-wideband transmission, enabling unprecedented data rates and supporting bandwidth-intensive applications like holographic telepresence and immersive virtual reality \cite{6g,6G_Dang,IY_MultiRIS_THz,SimRIS_Mag}. However, the migration to THz frequencies introduces fundamental challenges, particularly in wideband massive multiple-input multiple-output (MIMO) systems. 
%Among these, the beam split effect, caused by frequency-dependent beam misalignment, significantly degrades system performance by reducing array gain across subcarriers.
Hybrid beamforming has become a cornerstone in the implementation of massive MIMO systems, blending analog and digital beamforming to strike a balance between hardware complexity and system performance. However, in wideband systems, the \textit{beam split effect}—a phenomenon where the beams for different subcarriers become misaligned due to frequency-dependent phase shifts—poses a critical challenge \cite{HBF_THz}. This effect is particularly pronounced in wideband  THz and  mmWave systems, where frequency-dependent misalignments severely degrade beamforming efficiency \cite{HBF_Wideband}. 

Beam split mitigation has been extensively studied in wideband communication systems, particularly in the context of mmWave and THz frequencies. Several researchers have explored hardware-based solutions to mitigate the beam split effect in wideband THz massive MIMO systems. Traditional hybrid beamforming methods often employ narrowband approximations, which fail to address the beam misalignment in wideband scenarios \cite{Heath_OverviewMIMO,Heath_2014}. One approach involves employing multiple RF chains to generate frequency-dependent beamforming vectors, which can effectively alleviate beam squint across the wide bandwidth \cite{Survey_Wideband}. However, this method increases hardware complexity and cost due to the additional RF chains required. Alternatively, integrating true-time-delay (TTD) units into phase shifters has been proposed to compensate for the beam squint effect by aligning the phases of signals at different frequencies \cite{TTD_Hunter, TTD_THz}. While TTD units can improve beamforming accuracy over a wide frequency range, their implementation necessitates significant modifications to existing transceiver hardware, posing challenges for practical deployment.

Several studies have explored hybrid  TTD architectures to reduce the required range of time delays in wideband systems. The authors in \cite{wan2021hybrid} proposed a new hybrid precoding and combining scheme that forms a wide beam radiation pattern to mitigate the beam squint effect in mmWave/THz MIMO systems. However, their approach overlooks the beam split effect and does not consider uniform rectangular arrays (URAs) or multi-antenna user scenarios. Furthermore, a hybrid TTD architecture is proposed in \cite{boljanovic2021fast} for rapid beam training, combining analog time delays with a small number of digital time delays. While this approach mitigates some hardware constraints, the reliance on delays matching the number of antennas in existing works \cite{boljanovic2021fast,hashemi2008integrated} results in high power consumption, especially during precoding with multiple data streams. Additionally, these TTD-based architectures primarily focus on beamformer design for a single data stream, leaving the challenges of precoding design for multiple data streams unaddressed. In  \cite{DPP_Full}, the authors propose a novel hybrid precoding architecture, termed delay-phase precoding (DPP), which incorporates a time-delay network into traditional analog beamforming. This design enables frequency-dependent beamforming,  mitigating the beam split effect and enhancing the achievable performance in wideband THz massive MIMO systems. However, these methods require additional hardware, increasing system cost and complexity, and limiting scalability.
\begin{figure*}[!t]
	\centering
	\includegraphics[width=1.25\columnwidth]{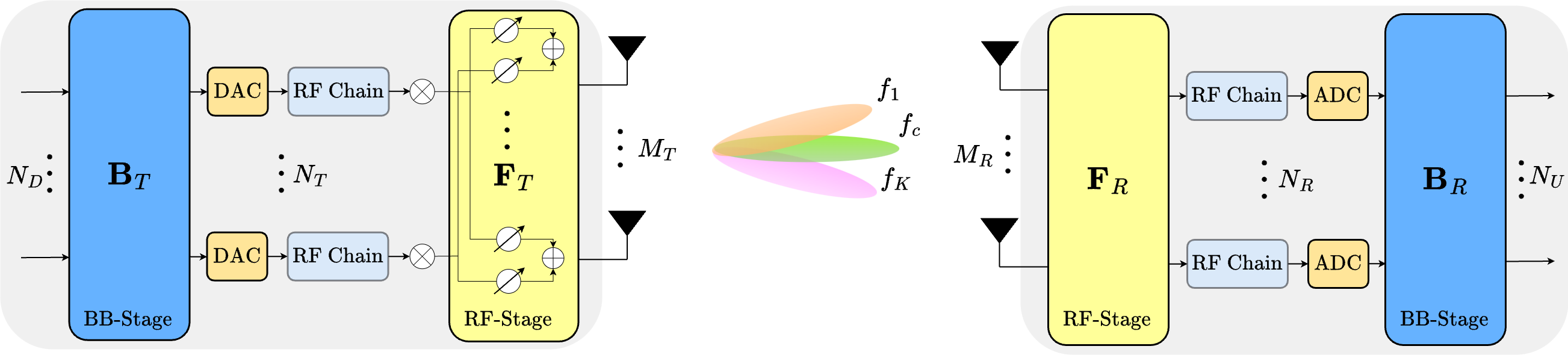}
	\vspace{-2ex}
	\caption{System model of AB-HBF wideband massive MIMO.}
	\vspace{-4ex}
	\label{fig:system} 
\end{figure*}	

Recent studies have explored angular-based hybrid beamforming techniques (AB-HBF) to reduce hardware complexity and channel state information (CSI) overhead in mmWave/THz massive MIMO systems \cite{RIS_AB_HPC,ASIL_ABHP_Access,SpiderRIS}. In \cite{ASIL_ABHP_Access}, a hybrid mmWave massive MIMO system with low-resolution digital-to-analog converters (DACs) and analog-to-digital converters (ADCs), along with reduced CSI overhead, is introduced.
 Analog beamformers are designed using slow-varying angle-of-departure (AoD) and angle-of-arrival (AoA) parameters, while baseband precoders employ reduced-size effective CSI. This approach highlights the potential of leveraging angular information to achieve high spectral efficiency with lower hardware cost and minimal CSI requirements.

Building upon these concepts, we propose an angular-based hybrid beamforming framework for wideband THz massive MIMO systems that mitigate the beam split effect without the need for TTD units. By designing the analog beamformers using coarse angular information and dynamically adjusting the angular spread, our method effectively reduces beam misalignment across subcarriers inherent in wideband systems. The baseband and RF beamformers are jointly designed at the transmitter, providing a flexible and low-cost solution to the beam split problem. Moreover, our approach extends to scenarios involving multi-antenna users and URAs, offering enhanced adaptability and performance in practical THz communication environments.

The remainder of this work is organized as follows: Section II describes the channel and system model used in our analysis. In Section III, we delve into the beam split effect in wideband THz massive MIMO systems. Section IV presents the proposed AB-HBF design for wideband channels. Numerical results evaluating the performance of the proposed method are provided in Section V. Finally, Section VI concludes the work.

%Although substantial progress has been made in hybrid beamforming and beam split mitigation, existing methods remain constrained by their reliance on hardware-intensive solutions or computationally expensive channel modeling approaches. The integration of angular spread as a design parameter for mitigating beam split has been largely overlooked.

\section{System Model}

In this section, we introduce the channel and  signal model for the system depicted in Fig. \ref{fig:system}. We consider a wideband THz massive MIMO system with a URA at both the transmitter and receiver. The system operates over a wideband channel utilizing orthogonal frequency division multiplexing (OFDM) with $K$ subcarriers. The transmitter comprises $M_T = M_{T_x} \times M_{T_y}$ antenna elements, where $M_{T_x}$ and $M_{T_y}$ denote the number of rows and columns of the array, respectively. Similarly, the receiver is equipped with $M_R = M_{R_x} \times M_{R_y}$ antenna elements arranged in a URA configuration analogous to the transmitter. The transmit baseband precoder and RF beamformer are interconnected through $N_T$ RF chains, while the receiver has $N_R$ RF chains to facilitate the connection.
 In practical implementations of massive MIMO systems with large-scale antenna arrays, the number of RF chains is significantly smaller than the total number of antennas, i.e., $N_T \ll M_T$ and $N_R \ll M_R$. Communication between the transmitter and receiver is established through $N_D$ data streams, where $N_D \le N_T$. Therefore, the pre-processed received signal at the receiver for $k$-th subcarrier is expressed as
\begin{equation}
\mathbf{y}_k = \sqrt{\rho} \mathbf{H}_k \mathbf{F}_T^k \mathbf{B}_T^k  \mathbf{d}_k + \mathbf{n}_k,
\end{equation}
where $\rho$ represents the transmit power, $\mathbf{H}_k \in \mathbb{C}^{M_R \times M_T}$ is the channel matrix, $\mathbf{F}_T^k \in \mathbb{C}^{M_T \times N_T}$ and $\mathbf{B}_T^k \in \mathbb{C}^{N_T \times N_D}$ denote the baseband precoding and RF beamformer matrices, $\mathbf{d}_k \in \mathbb{C}^{N_D \times 1}$ is the vector of transmitted data symbols, and $\mathbf{n}_k \in \mathbb{C}^{M_R \times 1}$ is the additive Gaussian noise vector with zero mean and variance $\sigma^2$. Furthermore, the received signal is first combined by an RF beamformer $\mathbf{F}_R^k \in \mathbb{C}^{N_R \times M_R}$ and then processed digitally via a combiner $\mathbf{B}_R^k \in \mathbb{C}^{N_U \times N_R}$ with $N_R$ receive RF chains where $N_U \le N_R \ll M_R $. For simplicity, any beam deviation at the receiver is neglected by assuming $M_R \ll M_T $.

%where $\rho$ represents the transmit power,$\mathbf{H}_k \in \mathbb{C}^{M_R \times M_T}$ denotes the frequency-selective channel matrix, $\mathbf{F}_k \in \mathbb{C}^{M_T \times N_T}$ and $\mathbf{B}_k \in \mathbb{C}^{N_T \times N_D}$ are the hybrid precoding and combining matrices, $\mathbf{d}_k \in \mathbb{C}^{N_D \times 1}$ is the transmitted symbol vector, and $\mathbf{n}_k \sim \mathcal{CN}(0, \sigma^2)$ is the additive Gaussian noise vector.

%In this section, we have introduced the channel  and beam signal model for the system given in Fig. \ref{fig:system}.
%We consider a wideband THz massive MIMO system with a URA at the transmitter and receiver. The system operates over a wideband channel using orthogonal frequency division multiplexing (OFDM) with $ K $  subcarriers. The transmitter consists of $M_T = M_{T_x} \times M_{T_y}$ antenna elements, where $M_{T_x}$ and $M_{T_y}$ are the number of rows and columns of the array, respectively. The receiver employs $M_R= M_{R_x} \times M_{R_y}$ antenna elements arranged in URA  similar to the transmitter. 

 The channel matrix for the $k$-th subcarrier  $
\mathbf{H}_k \in \mathbb{C}^{M_R \times M_T} $
is modeled using a geometric channel model as \cite{THz_channel_model, BeamSquint_THz}
\begin{equation} \label{eq:channel}
\mathbf{H}_k =  \sum_{\ell=1}^L \alpha_\ell(f_k,d_T) e^{-j2\pi \tau_\ell f_k}\mathbf{a}_{\text{R}}(\theta_\ell^r, \psi_\ell^r,f_k) \mathbf{a}_{\text{T}}^H(\theta_\ell^t, \psi_\ell^t,f_k),
\end{equation}
where $L$ represents the number of propagation paths between the transmitter and receiver. The term $\alpha_\ell(f_k, d_T)$ is the complex gain associated with the $\ell$-th path, and it is given by
\begin{equation}
    \alpha_\ell(f_k, d_T) = \frac{c}{4\pi f_k d_T} e^{-\frac{1}{2} \tau_a(f_k) d_T},
\end{equation}
where $c$ is the speed of light, $f_k$ is the subcarrier frequency, and $d_T$ is the distance between the transmitter and receiver. The factor $e^{-\frac{1}{2} \tau_a(f_k) d_T}$ accounts for the molecular absorption effect at the frequency $f_k$.

The array response vector at the receiver, $\mathbf{a}_{\text{R}}(\theta_\ell^r, \psi_\ell^r, f_k)$, depends on the AoA of the $\ell$-th path, defined by the azimuth angle $\theta_\ell^r$ and elevation angle $\psi_\ell^r$. Similarly, the array response vector at the transmitter, $\mathbf{a}_{\text{T}}(\theta_\ell^t, \psi_\ell^t, f_k)$, depends on the AoD of the $\ell$-th path, characterized by the azimuth angle $\theta_\ell^t$ and elevation angle $\psi_\ell^t$.
For $i \in \{\text{T}, \text{R}\}$, the transmit/receive array responses are given by:
\begin{equation}\label{eq:array_response}
	\begin{split}
		\mathbf{a}_i\left( \theta_\ell^i, \psi_\ell^i, f_k \right) &= \big[ 1, \cdots, e^{j\pi  \frac{f_k}{f_c} \left( M_{i_x} - 1 \right) \sin (\theta_\ell^i) \cos (\psi_\ell^i)} \big]^T \\
		& \otimes \big[ 1, \cdots, e^{j\pi  \frac{f_k}{f_c} \left( M_{i_y} - 1 \right) \sin (\theta_\ell^i) \sin (\psi_\ell^i)} \big]^T,
	\end{split} \vspace{-0.5ex} \raisetag{0.8\baselineskip}
\end{equation}
where $f_c$ is the carrier frequency, and $M_{i_x}$ and $M_{i_y}$ denote the number of antennas in the row  and column directions for $i \in \{\text{T}, \text{R}\}$, respectively. This formulation captures the spatial and frequency-dependent phase shifts of the antenna array.

\section{Beam Split Analysis}
In wideband THz massive MIMO systems, the beam split phenomenon primarily stems from the frequency-dependent nature of the spatial array responses. As captured in \eqref{eq:array_response}, the effective steering directions depend on both the physical propagation angles and the operating frequency. When the carrier frequency 
$f_c$ is scaled to different subcarrier frequencies 
$f_k$ ($k=1,\dots,K$), the resulting array phase increments become frequency-dependent. This causes the beam steering direction at each subcarrier to diverge from its intended target, producing what is commonly referred to as \textit{beam split}.

This issue is further aggravated by several factors intrinsic to THz communications. First, the exceptionally wide bandwidths at THz frequencies intensify even minor frequency-induced deviations, substantially altering the beam directions across the spectrum. Second, the large-scale antenna arrays employed in THz massive MIMO systems generate extremely narrow beamwidths. Under such circumstances, even slight frequency offsets between subcarriers can lead to significant misalignment, thereby reducing the effective array gain. Lastly, other frequency-dependent effects—such as the path-dependent time delays from propagation (as in \eqref{eq:channel}) and frequency-selective molecular absorption—compound the beam split effect. In essence, the combination of large bandwidth, massive arrays, and frequency-sensitive propagation parameters renders beam split a critical, and increasingly pronounced, challenge in THz massive MIMO deployments.

The wideband nature of the channel includes the frequency-dependent phase shifts due to both the propagation paths and the spatial positioning of the antenna elements. The term \(e^{-j2\pi \tau_\ell f_k}\) in \eqref{eq:channel} introduces the frequency-dependent time delays, which are essential in wideband systems to account for the beam split effect that occurs across different subcarriers.
This misalignment in beams leads to a reduction in the effective array gain and overall system performance, especially in THz wideband systems.

To characterize this effect mathematically, we consider a particular propagation path \(\ell\) with azimuth angle \(\theta_\ell\) and elevation angle \(\psi_\ell\) defined at the carrier frequency \( f_c \). For the \( k \)-th subcarrier at frequency \( f_k \), we define:
\begin{equation}\label{eq:ThetaPsi_def}
\Theta_{k,\ell} = \frac{f_k}{f_c} \sin(\theta_\ell) \cos(\psi_\ell), \quad 
\Psi_{k,\ell} = \frac{f_k}{f_c} \sin(\theta_\ell) \sin(\psi_\ell).
\end{equation}
In the narrowband case (i.e., when \( f_k 	\approx f_c \)), the target angular parameters reduce to
$
\Theta = \sin(\theta_\ell)\cos(\psi_\ell)$ and \ $\Psi = \sin(\theta_\ell)\sin(\psi_\ell)$. To measure the frequency-induced deviations, we define the physical direction components as
\begin{equation}\label{eq:DeltaThetaPsi_def}
\Delta \Theta_k = \Theta_{k,\ell} - \Theta, \quad \Delta \Psi_k = \Psi_{k,\ell} - \Psi.
\end{equation}
%These deviations \(\Delta \Theta_k\) and \(\Delta \Psi_k\) quantify how much the actual steering direction for subcarrier \(k\) differs from the intended steering direction at \( f_c \).
The beam misalignment caused by \(\Delta \Theta_k\) and \(\Delta \Psi_k\) leads to a reduction in the effective array gain. Here, a single-antenna receiver is considered for simplicity, although extending the analysis to multiple antennas is straightforward. For a URA with \( M_{T_x} \) and \( M_{T_y} \) antennas in the row and column dimensions with $ M_T = M_{T_x} \times M_{T_y}$, the normalized array gain at the $k$-th subcarrier can be written as:
\begin{align}\label{eq:eta_k_def}
\eta_k(\mathbf{f}_{\ell},\theta_\ell^t, \psi_\ell^t,f_k) =&\left| \mathbf{a}_{\text{T}}^H(\theta_\ell^t, \psi_\ell^t,f_k)\mathbf{f}_{\ell} \right| \nonumber \\
=\frac{1}{M_{T_x} M_{T_y}}& \left| \sum_{x=0}^{M_{T_x}-1} \sum_{y=0}^{M_{T_y}-1} 
\exp\bigl(j \pi (x \Delta \Theta_k + y \Delta \Psi_k)\bigr) \right|.
\end{align}
%\mathbf{a}_{\text{T}}^H(\theta_\ell^t, \psi_\ell^t,f_k)
Approximating the summation by products of Dirichlet sinc functions, we have:
\begin{equation}\label{eq:eta_k_sinc}
\eta_k(\mathbf{f}_{\ell},\theta_\ell^t, \psi_\ell^t,f_k) \approx \left| \text{sinc}\left(\frac{M_{T_x}\pi \Delta \Theta_k}{2}\right) 
\text{sinc}\left(\frac{M_{T_y}\pi \Delta \Psi_k}{2}\right) \right|,
\end{equation}
where \(\text{sinc}(x) = \sin(x)/x\).
As the bandwidth increases, the ratio \( \frac{f_k}{f_c} \) diverges more substantially from unity, leading to larger values of \(\Delta \Theta_k\) and \(\Delta \Psi_k\). Consequently, the array gain \(\eta_k\) degrades across subcarriers, making beam split a prominent issue that must be addressed in wideband THz massive MIMO system designs.
%To quantify the severity of the beam split effect, the Beam Split Ratio (BSR) is introduced. The BSR for UPA is defined as:
%\begin{equation}
%\text{BSR} = \frac{1}{M_T} \int_{-\pi/2}^{\pi/2} \int_{-\pi}^{\pi} \sum_{k=1}^K \frac{|f_k - f_c|}{f_c}  \left(\Delta \Theta_k + \Delta \Psi_k\right) d\theta d\psi.
%\end{equation}
%Higher values of BSR indicate severe beam misalignment and significant array gain degradation. The beam split effect becomes more pronounced with increasing carrier frequency $f_c$, larger bandwidth $B$, and greater array dimensions $M_{T_x}$ and $M_{T_y}$.

%The hybrid beamforming approach used in this system is designed to mitigate the beam split by aligning the beams at the RF stage and optimizing the digital precoding stage. 
%In wideband systems, the frequency-dependent nature of the channel means that the alignment of the beams needs to be adjusted for each subcarrier to maintain consistent performance across the entire bandwidth. The hybrid beamforming framework, which combines analog beamforming at the RF stage with digital beamforming at the baseband stage, provides a scalable solution for mitigating the beam split effect. By exploiting the spatial characteristics of the channel, this approach reduces the beam misalignment across subcarriers and improves the overall signal quality.

\section{Angular-Based Hybrid Beamforming Design}

In this section, we present the AB-HBF design for wideband THz massive MIMO systems. The goal is to reduce hardware complexity and CSI overhead while mitigating the beam split effect. The AB-HBF design is based on leveraging slow-time varying (angular) information of the channel, specifically the  AoD and AoA distributions, to construct efficient RF and baseband precoders.

\subsection{RF-Stage Design}
\label{sec:rf_stage}

The key idea in the RF-stage design is to represent the spatial domain via quantized angular supports and then construct an RF beamformer that can efficiently cover the set of relevant AoDs. Let us focus on the transmitter side, where we aim to form a set of steering vectors that capture the dominant propagation directions. By carefully selecting these quantized angular directions, we ensure that the resulting RF beamformer provides sufficient angular coverage for all targeted paths while minimizing the number of active RF chains.

To define the AoD support for the $\ell$-th propagation path, we consider the elevation and azimuth angle sets, which incorporate both the nominal path angles and their corresponding angular spreads:
\begin{align} \label{eq:AoD}
    \Upsilon &= \big\{ [ \Theta, \Psi ] = \sin(\theta_\ell^t)[\cos(\psi_\ell^t), \sin(\psi_\ell^t)] \mid \theta_\ell^t \in \bm{\theta}_t, \;\psi_\ell^t \in \bm{\psi}_t \big\},
\end{align}
where 
$\bm{\theta}_t= \bigcup_{\ell=1}^{L} \left[\theta_{l}^t - \delta_{\theta_{t}}^{\ell},\;\theta_{l} + \delta_{\theta_{t}}^{\ell}\right]$ and $
\bm{\psi}_t= \bigcup_{\ell=1}^{L} \left[\psi_{l}^t - \delta_{\psi_{t}}^{\ell},\;\psi_{l} + \delta_{\psi_{t}}^{\ell}\right]$
 stand for the elevation and azimuth angle supports, respectively. Here, $\delta_{\theta_{t}}^{\ell}$ and $\delta_{\psi_{t}}^{\ell}$ characterize the angular spread around the central AoD angles, reflecting practical propagation conditions such as diffuse scattering and imperfect beam alignment.

By discretizing the angular domain into quantized angle-pairs, we can approximate the continuous angular space with a finite and manageable set of steering vectors. These quantized angle-pairs are chosen to ensure orthogonality and full coverage of the AoD support, allowing the RF beamformer to be constructed from a well-defined dictionary of possible steering directions as
\begin{equation}\label{eq:RFvector}
	{{\bf{F}}}_T^k = \big[ \mathbf{a}_{\text{T}} (\lambda_\theta^{x_1}, \lambda_\psi^{y_1},f_k) ,  \cdots ,  \mathbf{a}_{\text{T}} (\lambda_\theta^{x_{N_T}}, \lambda_\psi^{y_{N_T}},f_k)   \big],
\end{equation}
where the quantized angles are defined as
$
\lambda_\theta^{x}=\frac{2x-1}{M_{T_x}}-1$, for $x=1,\dots,M_{T_x}$
and $\lambda_\psi^{y}=\frac{2y-1}{M_{T_y}}-1$ for $ y=1,\dots,M_{T_y}$. These quantized angle-pairs form a complete set of $M_T = M_{T_x}\times M_{T_y}$ orthogonal steering vectors that can represent any direction within the considered angular sector. By selecting only those angle-pairs that lie within the defined AoD support \eqref{eq:AoD}, we can realize an RF-stage design that covers all relevant spatial directions without resorting to an excessive number of RF chains or incurring substantial CSI overhead.

\begin{figure}[!t]  
    \centering 
    \vspace*{0.04in}
    \subfigure[]{
    \includegraphics[height=4.96cm]{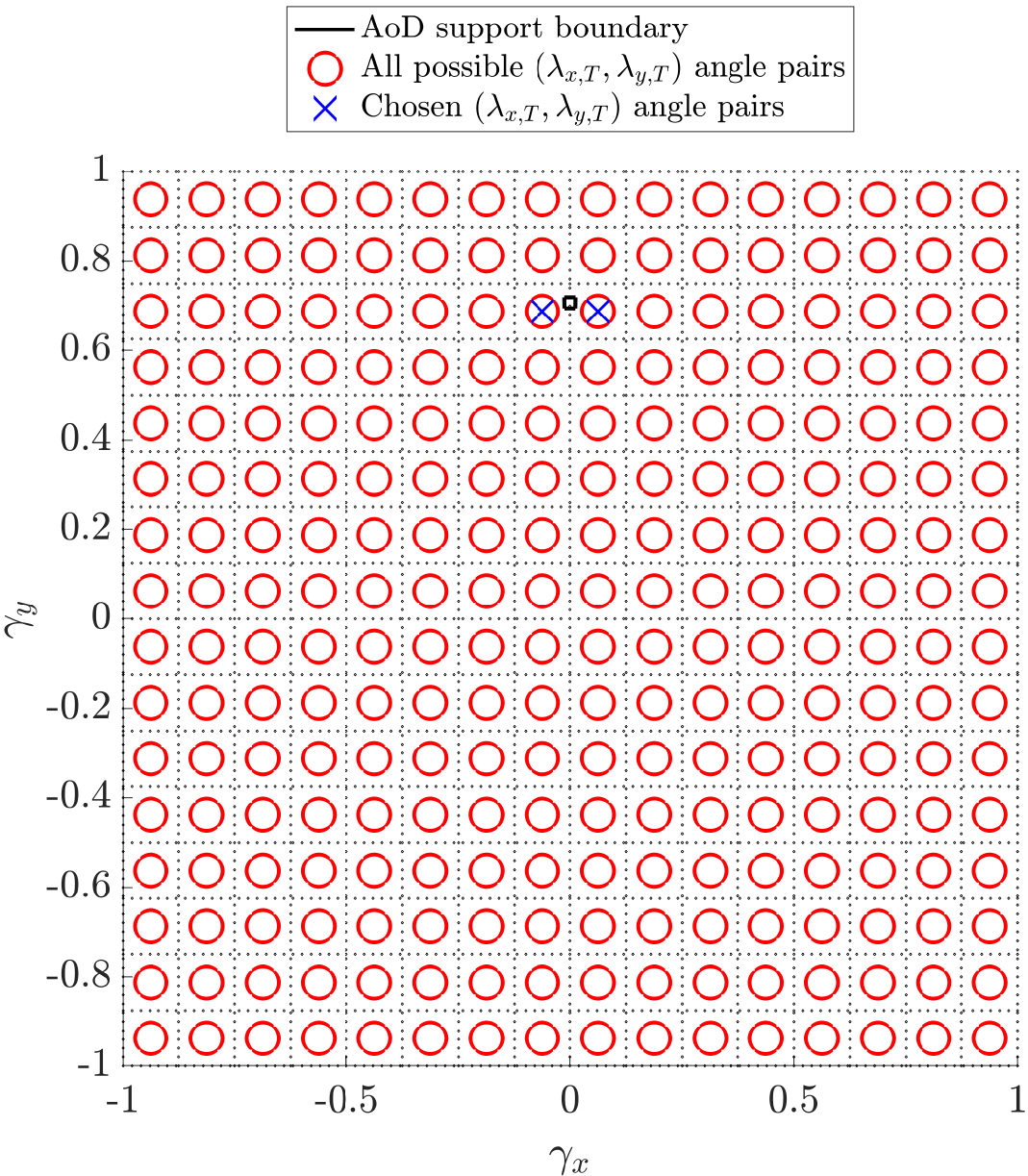}}
    \hspace{-0.3cm}  % Figürler arasındaki yatay boşluğu azaltmak için
    % İkinci figür
    %\subfigure[]{\includegraphics[height=7cm]{Figs/Fig_Ang_Sup_2.pdf}}
    %\hspace{-0.05\textwidth}  
    \subfigure[]{ \includegraphics[height=4.96cm]{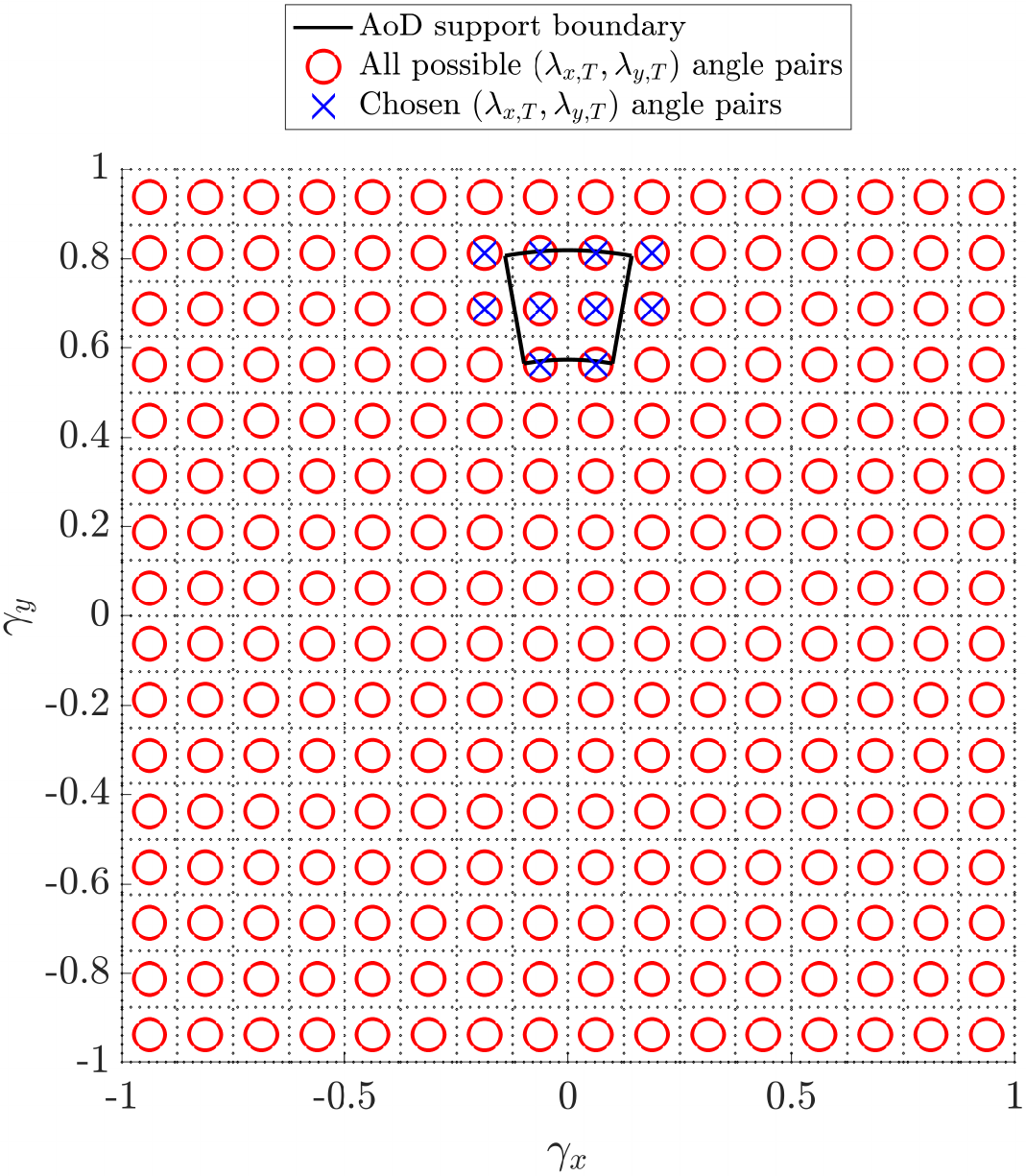}}\vspace{-1.6ex}
    \caption{RF beamformer design for a $16 \times 16$ URA for different angular spreads of (a) 2°  and (b) 10°.}
    \label{fig:angular_support} \vspace{-0.5cm}  
\end{figure}

Fig. \ref{fig:angular_support} illustrates the selection of quantized angle pairs for the RF beamformer design under varying angular spreads. Here, the red circles represent all possible angle pairs that span the entire angular domain for a $16 \times 16$ URA, and the solid black line outlines the AoD support boundary. The blue crosses mark the chosen angle pairs that lie within this support.
In the case of an angular spread 2$^\circ$, the AoD support boundary is more localized, resulting in a compact set of selected angle pairs. By contrast, when the angular spread increases to 10$^\circ$, the AoD support boundary encompasses a wider angular region, prompting the selection of more quantized angle pairs. This expansion enables coverage of a broader spatial domain, ensuring that the RF beamformer remains robust against increased angular dispersion.

%In order to use the flexibility and scalability of our proposed AB-HBF framework to mitigate beam split, we investigate the effect of varying angular spread on normalized array gain in Fig. \ref{fig:ang_spread_sub} for different subcarriers.

\begin{figure}[!t]
	\centering
    \vspace*{0.04in}
	\includegraphics[width=0.7\columnwidth]{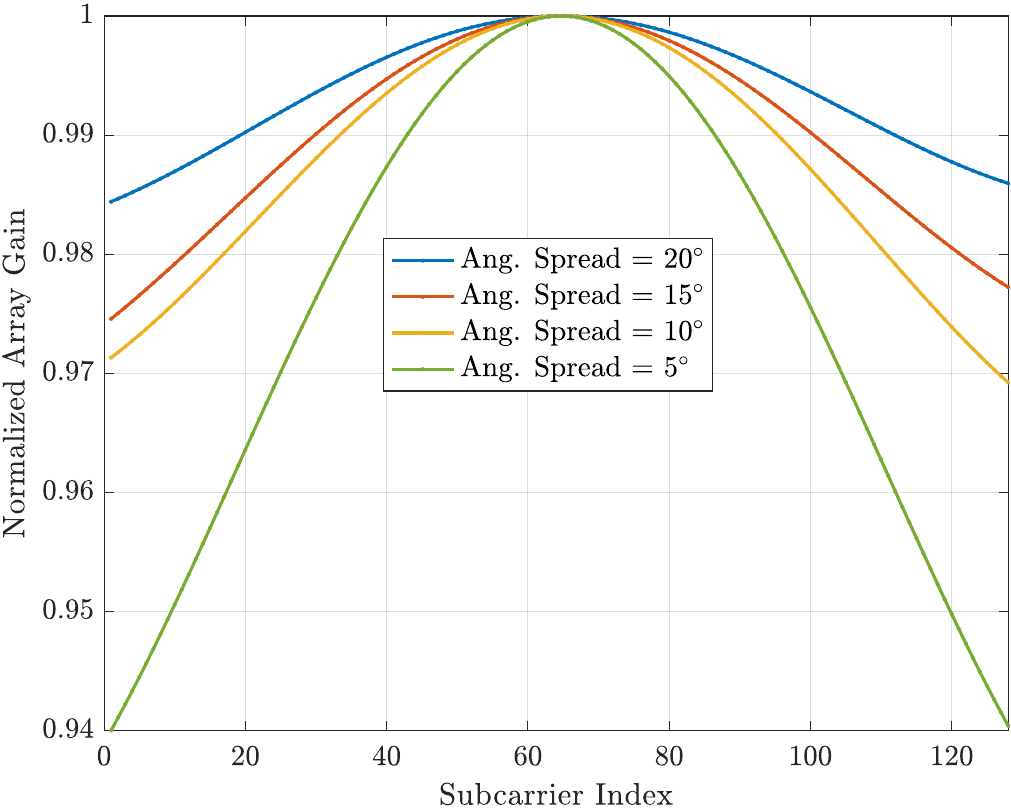}
	\vspace{-2ex}
	\caption{Normalized array gain of the RF beamformer for varying angular spreads.}
	\vspace{-3ex}
	\label{fig:ang_spread_sub} 
\end{figure}

Fig. \ref{fig:ang_spread_sub} demonstrates the influence of different angular spreads on the normalized array gain across multiple subcarriers. Larger angular spreads result in a more resilient array gain profile, indicating that the flexible AB-HBF framework can better maintain beam coherence over the wideband spectrum. By strategically selecting angular supports and quantized angle pairs, the AB-HBF approach effectively mitigates the beam split effect, ensuring a more uniform and stable array gain distribution even as the system bandwidth and the spatial dimensions grow.

%This comparison demonstrates the flexibility and scalability of our proposed AB-HBF framework. 
%As the angular spread grows, our quantized angle-pair selection strategy adaptively includes additional steering vectors, maintaining effective beam coverage while minimizing hardware complexity and CSI requirements. Such adaptability is crucial in wideband THz massive MIMO systems, where dynamic channel conditions can rapidly alter the spatial characteristics of the propagation environment.

%As the angular spread increases, the beamforming pattern becomes more spread, resulting in reduced beam split and improved alignment across subcarriers. The figure demonstrates how increasing the angular spread enhances the system's performance by mitigating the frequency-dependent misalignment of beams.

% In doing so, the AB-HBF framework effectively balances the conflicting requirements of hardware complexity and spatial coverage, serving as a foundation for the subsequent baseband beamforming and providing a robust means of mitigating the beam split effect in wideband THz massive MIMO systems.

\subsection{Baseband Precoder Design}
\label{sec:bb_precoder}
After constructing the RF beamformers, the effective channel for $k$th subcarrier is defined as  $
    \mathbf{H}_\text{eff}^k =  \mathbf{H}_k \mathbf{F}_{T}^k$.
The baseband precoder is designed to maximize the effective channel gain using singular value decomposition (SVD). The SVD of $\mathbf{H}_\text{eff}$ is calculated as
\begin{equation}
    \mathbf{H}_\text{eff}^k  = \mathbf{U} \mathbf{\Sigma} \mathbf{V}^H,
\end{equation}
where $\mathbf{U}$ and $\mathbf{V}$ are unitary matrices, and $\mathbf{\Sigma}$ is a diagonal matrix containing the singular values. The baseband precoder is constructed as:
\begin{equation}
    \mathbf{B}_{T}^k  = \mathbf{V}_1 \mathbf{P},
\end{equation}
where $\mathbf{V}_1$ contains the first $N_D$ columns of $\mathbf{V}$, and $\mathbf{P}$ is a diagonal matrix satisfying the transmit power constraint.

The proposed AB-HBF framework efficiently designs the RF and baseband precoders using the angular properties of the channel. By leveraging angular sparsity, the framework reduces the hardware complexity and mitigates the beam split effect in wideband THz systems.

%Figure \ref{fig:angular_support} illustrates the impact of angular spread on mitigating the beam split effect for a  Uniform Rectangular Array (URA)  with $M_T = 256$ antennas. In this analysis, angular spread values of  2°, 5°, and 10°  are considered to observe how increasing angular spread affects the beam split across the wideband spectrum.

%  Beam Split Effect : Beam split occurs when different subcarriers experience misalignment due to frequency-dependent phase shifts. This causes a reduction in array gain and spectral efficiency. For narrower angular spreads (e.g., 2°), the beams are relatively narrow, and the beam split effect is more pronounced, resulting in greater performance degradation.

  %Mitigation of Beam Split : As the angular spread increases, the beams broaden, which reduces the beam split effect. Specifically, with an angular spread of 10°, the beams become sufficiently wide to align better across the subcarriers, minimizing misalignment and improving system performance. 

%This figure demonstrates that increasing the angular spread (e.g., from 2° to 10°) significantly mitigates the  beam split effect, ensuring that the beams from the transmitter align better at all subcarriers. As a result, the system performance improves, particularly in wideband THz communication systems with large antenna arrays.
%The results emphasize the importance of considering angular spread in the design of Angular-Based Hybrid Beamforming (AB-HBF) systems to optimize the performance of massive MIMO systems, especially in wideband scenarios where beam split is a key challenge.

\section{Numerical Results}
In this section, we present the simulation results to evaluate the performance of the proposed AB-HBF scheme in wideband THz massive MIMO systems. The simulation setup is configured with a carrier frequency of \( f_c = 300 \) GHz and an operating bandwidth of $30$ GHz with \( K = 128 \) subcarriers. The remaining system parameters are as follows unless otherwise stated:  $ d_T = 10$ m, \( M_T = 1024 \),  \( L = 4 \), $\delta_\psi = \delta_\theta = 10^\circ$ and $M_R=N_D=4$.

\begin{figure}[htbp]  
    \centering
    \subfigure[]{
    \includegraphics[height=5.2cm]{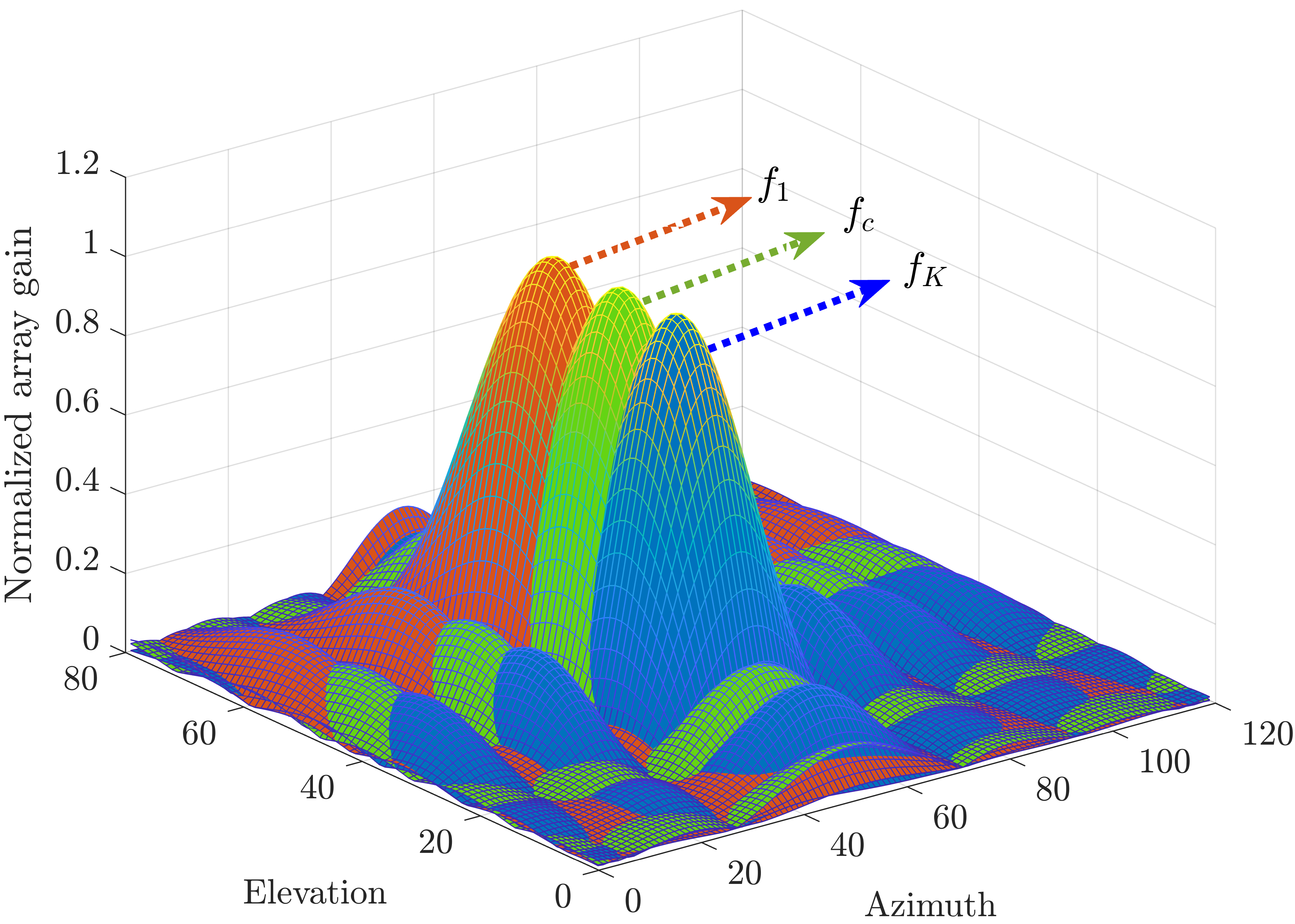}}
    \hspace{-0.25cm}  % Figürler arasındaki yatay boşluğu azaltmak için
    % İkinci figür
    \subfigure[]{\includegraphics[height=5.2cm]{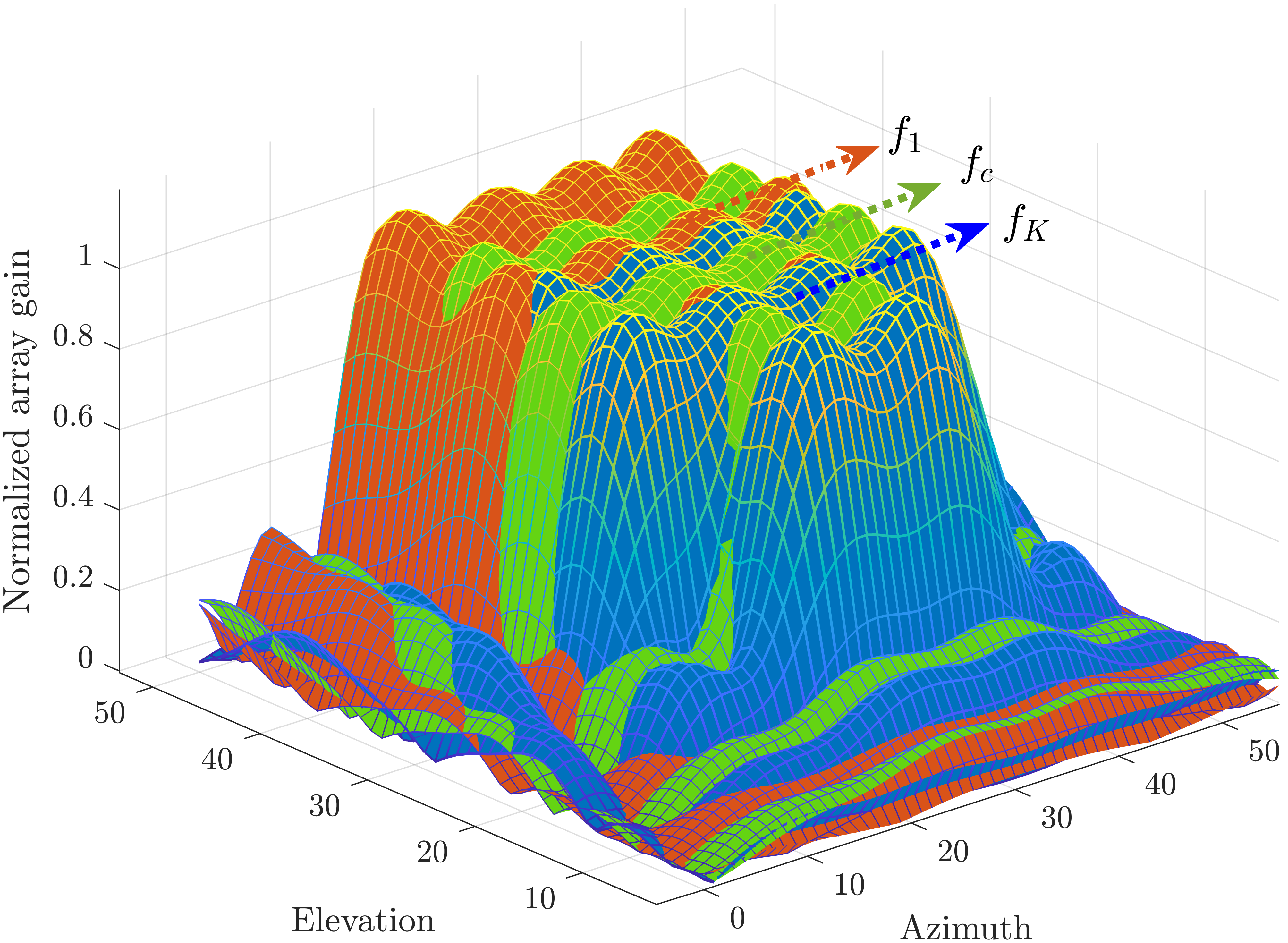}}
    \vspace{-0.2cm}  
    \caption{Comparison of normalized array gain between (a) narrowband conventional beamformer and (b) AB-HBF in a wideband THz massive MIMO system.}
    \label{fig:3D_gain} \vspace{-0.4cm} 
\end{figure}

\begin{figure}[!t]
	\centering
	\includegraphics[width=0.9\columnwidth]{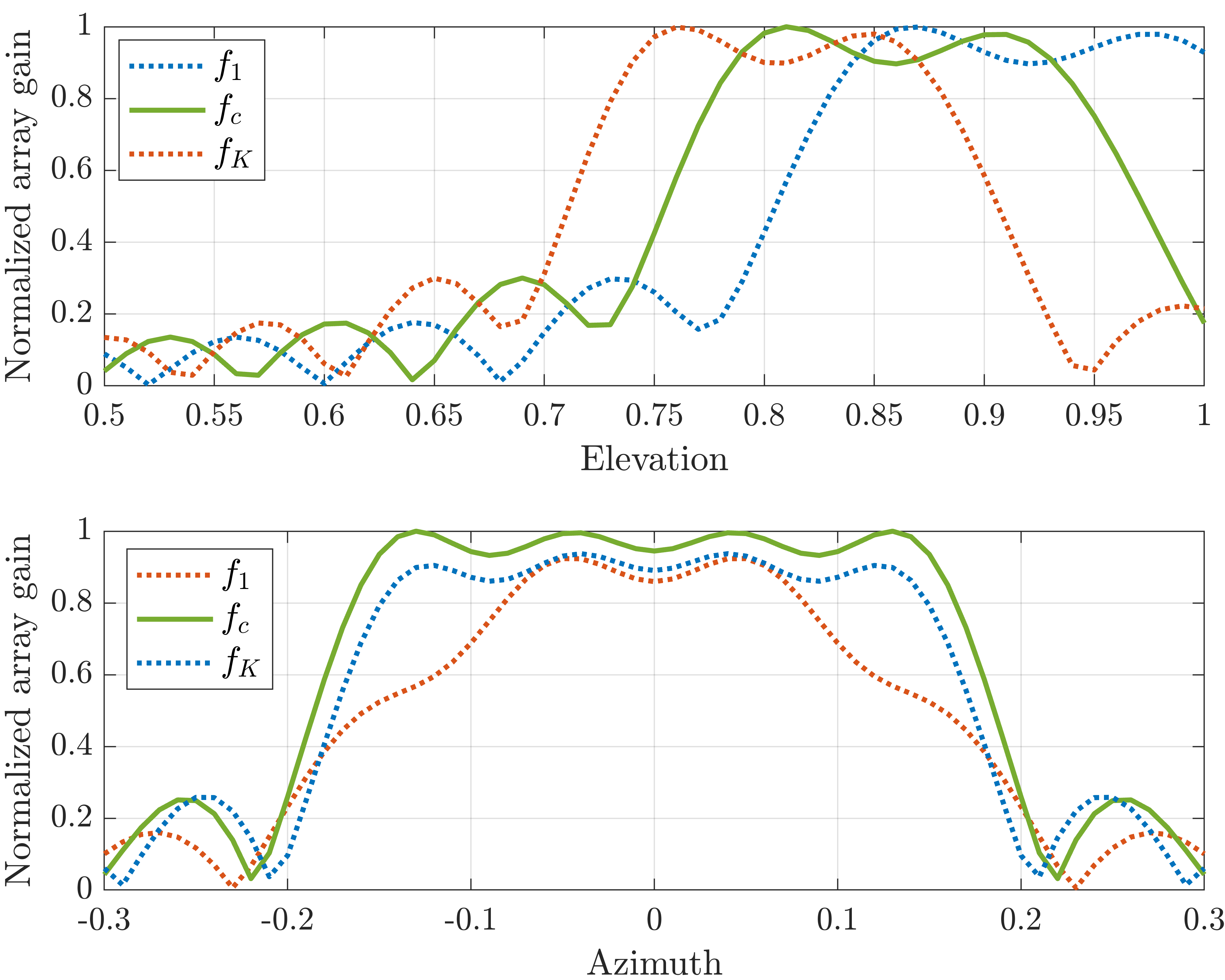}
	
	\caption{Normalized array gain of the proposed AB-HBF scheme versus elevation and azimuth angles in a wideband THz massive MIMO system.}
	\vspace{-3ex}
	\label{fig:Az_El} 
\end{figure}
Fig.~\ref{fig:3D_gain} compares the normalized array gain achieved by a conventional narrowband angular beamformer and the proposed AB-HBF technique in a wideband THz massive MIMO system. Fig.~\ref{fig:3D_gain}(a) illustrates the array gain distribution for a narrowband angular beamformer, where significant beam split occurs across different subcarriers, leading to reduced and inconsistent array gains. In contrast, Fig.~\ref{fig:3D_gain}(b) presents the 3D array gain achieved by AB-HBF, demonstrating a more uniform and higher array gain across the entire frequency band. This improvement is due to the AB-HBF's ability to dynamically adjust the angular spread and efficiently cover the spatial domain without relying on TTD units. Consequently, AB-HBF effectively mitigates the beam split effect, enhancing the normalized array gain and ensuring robust performance in wideband THz massive MIMO systems.

Figs.~\ref{fig:Az_El}(a) and (b) show the normalized array gain of the proposed AB-HBF scheme across varying elevation and azimuth angles, respectively, in a wideband THz massive MIMO system. In Fig.~\ref{fig:Az_El}(a), the AB-HBF maintains a normalized array gain close to $\eta_k=1$ across different elevation angles, demonstrating that the scheme effectively aligns the beams for all subcarriers near the central frequency with minimal deviation. Similarly, Fig.~\ref{fig:Az_El}(b) shows that the AB-HBF scheme achieves a normalized array gain near unity across various azimuth angles, with little variation observed among the subcarriers. These results highlight that the normalized array gain remains approximately $\eta_k=1$ for all subcarriers, indicating that the AB-HBF design successfully mitigates the beam split effect.

\begin{figure}[!t]
	\centering
	\includegraphics[width=0.8\columnwidth]{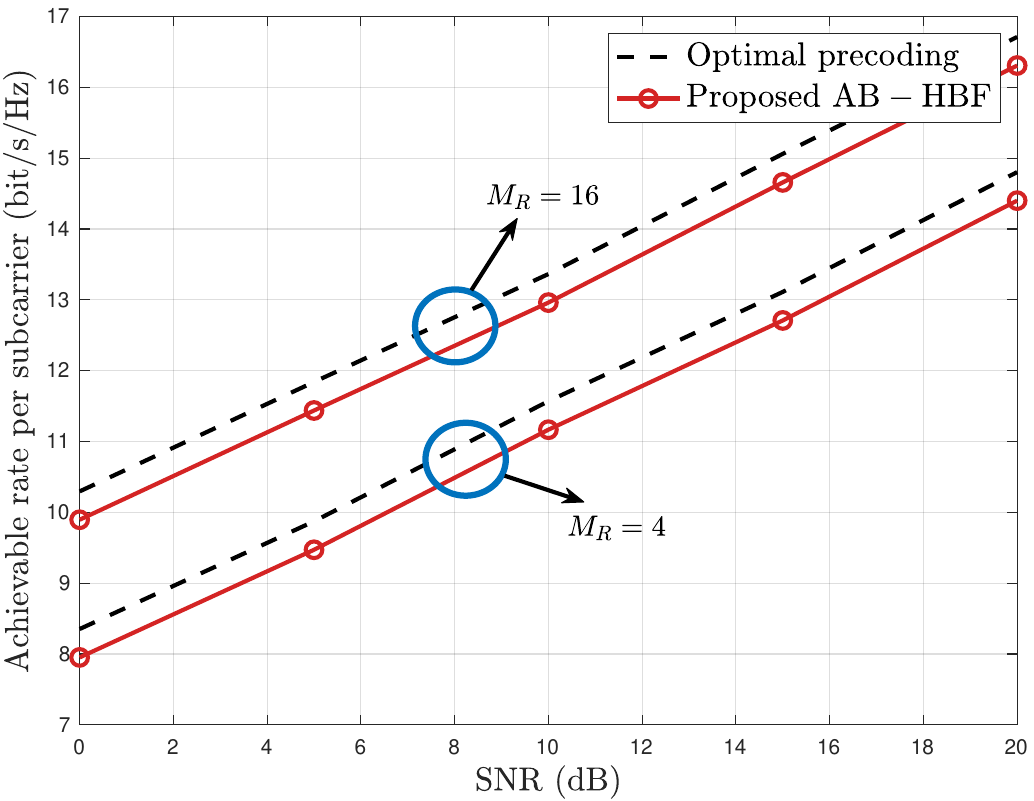}
	
	\caption{Achievable rate comparison between fully digital (optimal) precoding and the proposed AB-HBF scheme under varying SNRs in a wideband THz massive MIMO system for $M_T=4$ and $16$. }
	\vspace{-3ex}
	\label{fig:Rate_MR} 
\end{figure}
\begin{figure}[!t]
	\centering
	\includegraphics[width=0.8\columnwidth]{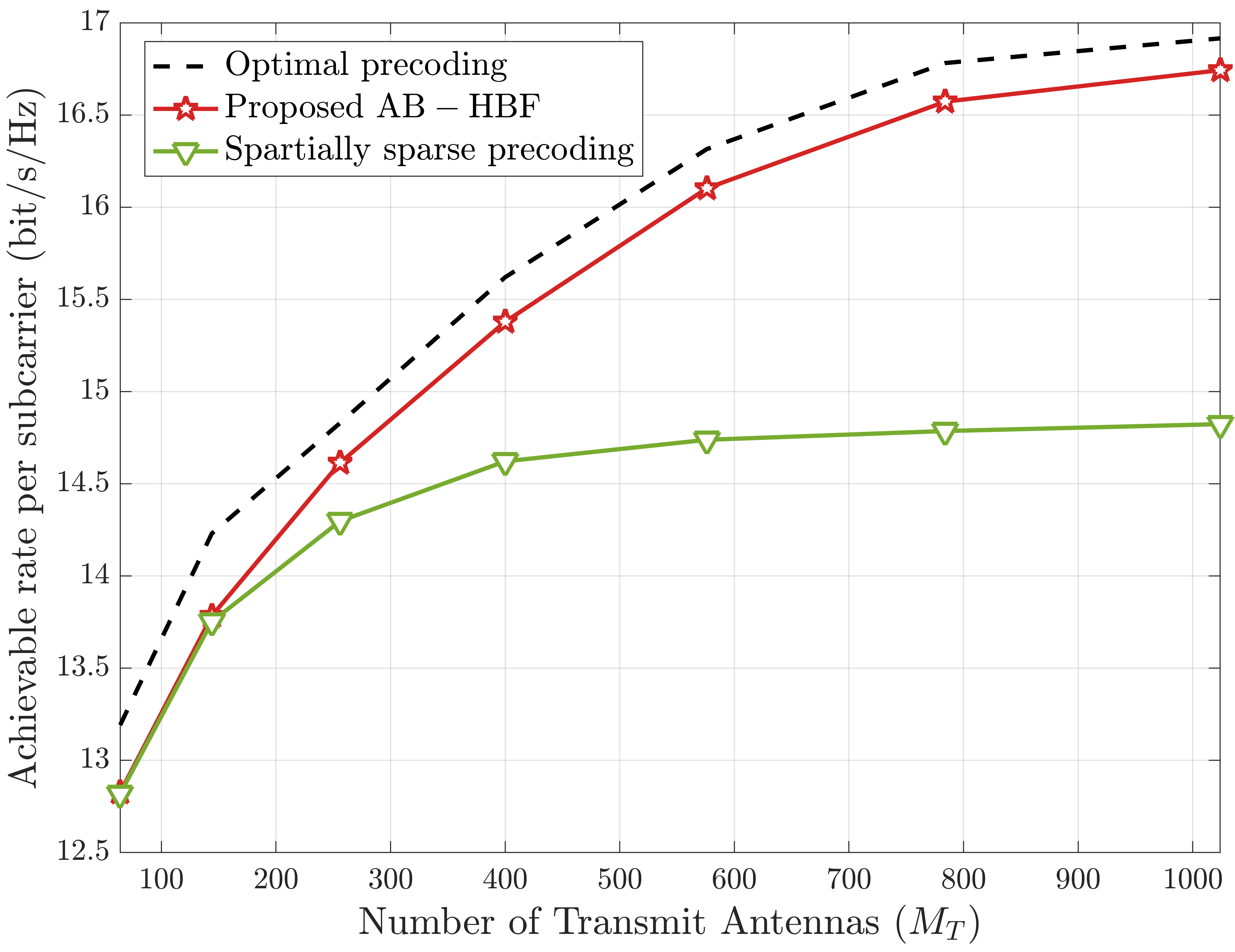}
	
	\caption{Achievable rates versus the number of Tx antenna elements of  AB-HBF, fully digital precoding, and traditional spatially sparse precoding.}
	\vspace{-3ex}
	\label{fig:Rate_MT} 
\end{figure}

Fig.~\ref{fig:Rate_MR} present the achievable rate of the proposed(AB-HBF scheme in comparison to fully digital precoding for increasing  SNR. The results demonstrate that AB-HBF achieves comparable performance to fully digital precoding while utilizing significantly fewer RF chains. Specifically, for varying numbers of receive antennas and data streams, AB-HBF maintains competitive achievable rates, highlighting its efficiency in reducing hardware complexity without sacrificing spectral efficiency. This performance is attributed to AB-HBF's effective use of angular information, which allows for optimized beamforming and substantial mitigation of the beam split effect. 
Furthermore, Fig.~\ref{fig:Rate_MT} compares the achievable rates of the AB-HBF scheme against fully digital optimal precoding and traditional narrowband spatially sparse precoding \cite{Heath_2014} under varying  $M_T$.  As the number of transmit antennas increases, the beam split effect in traditional narrowband precoding becomes more pronounced, leading to a substantial performance gap between AB-HBF and spatially sparse precoding method. This indicates that AB-HBF not only reduces hardware complexity but also ensures robust and reliable performance in wideband THz massive MIMO systems, making it a highly scalable and cost-effective solution.

\begin{figure}[!t]  
    \centering
    \subfigure[]{
    \includegraphics[height=5cm]{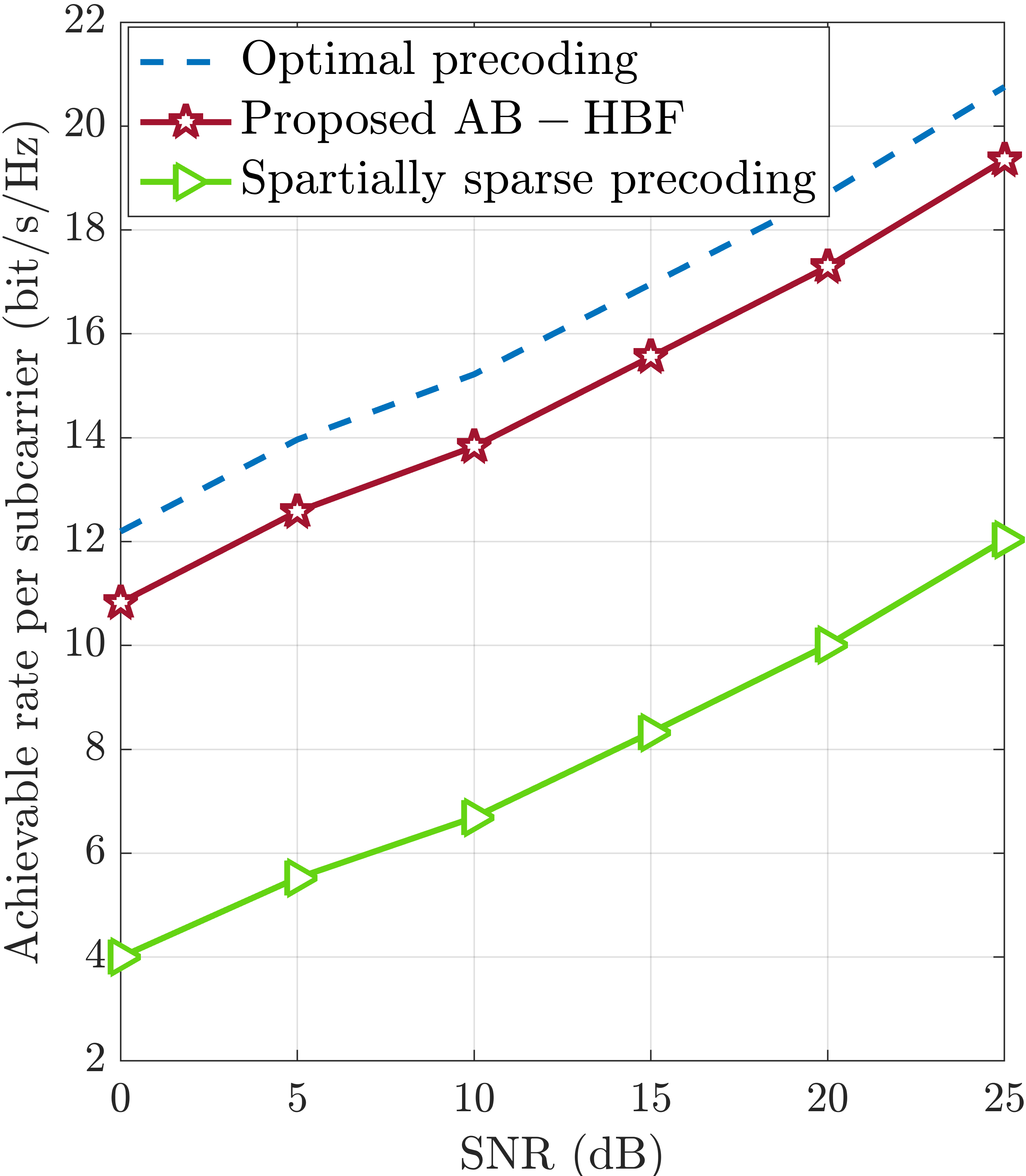}}
    \hspace{-0.25cm}  % Figürler arasındaki yatay boşluğu azaltmak için
    % İkinci figür
    \subfigure[]{\includegraphics[height=5cm]{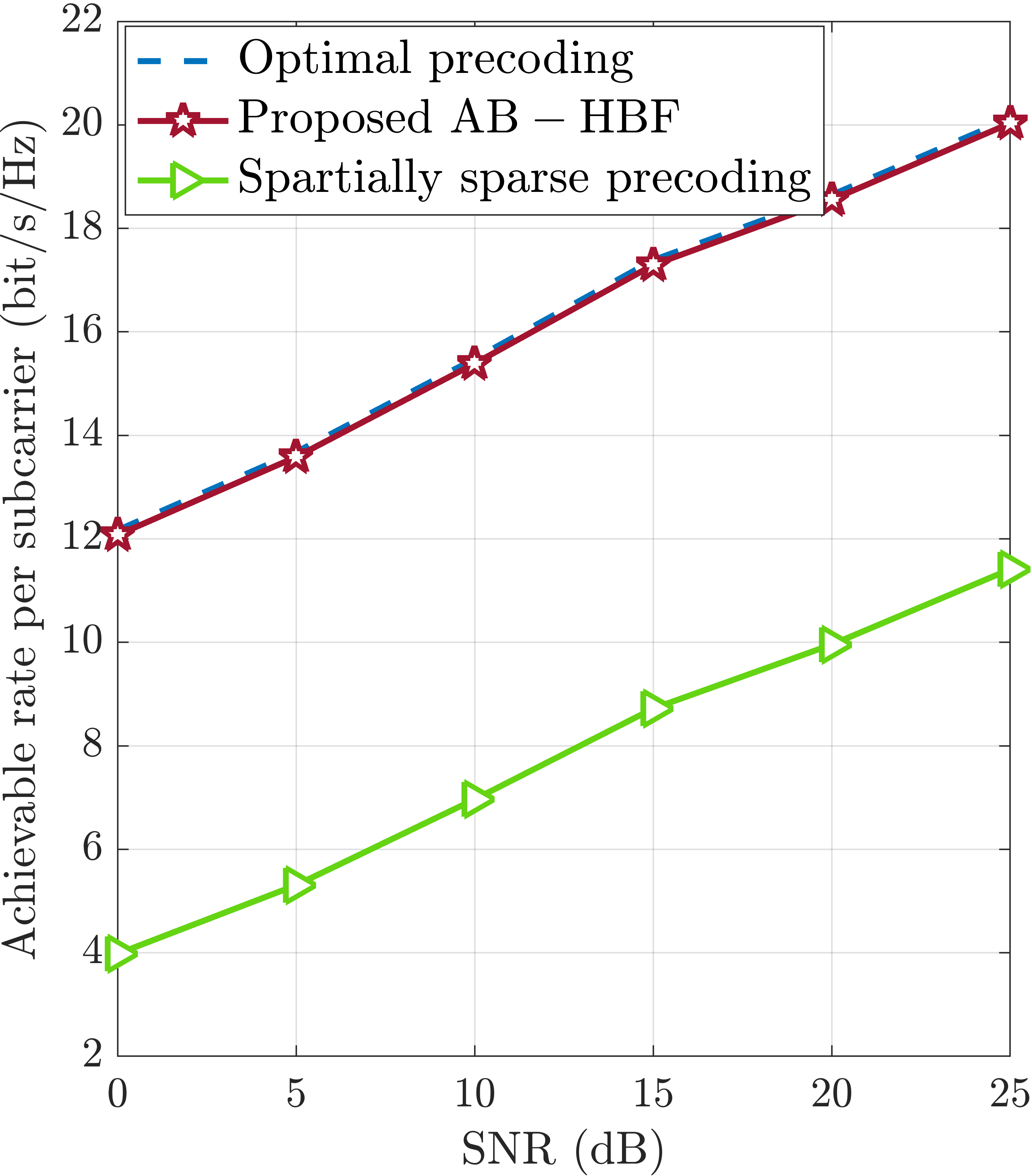}}
    \vspace{-0.4
    cm}  
    \caption{Achievable rates versus SNR  of  AB-HBF , fully digital precoding, and traditional spatially sparse precoding under  angular spreads of (a) $2^\circ$ and (b) $10^\circ$ in a wideband THz massive MIMO system.} 
    \label{fig:Rate_SNR}\vspace{-0.4cm} 
\end{figure}

Furthermore, Fig.~\ref{fig:Rate_SNR}(a) and (b) compare the achievable rates of the AB-HBF scheme against fully digital optimal precoding and traditional spatially sparse precoding \cite{Heath_2014} under varying angular spreads of 2$^\circ$ and 10$^\circ$, respectively. As observed from Fig.~\ref{fig:Rate_SNR}(a), even with a limited angular spread, AB-HBF effectively mitigates the beam split effect, leveraging angular information to enhance performance. In Fig.~\ref{fig:Rate_SNR}(b), as the angular spread increases to 10$^\circ$, the beam split effect is further reduced, resulting in the AB-HBF performance approaching that of fully digital precoding. 
These findings illustrate that the proposed AB-HBF scheme consistently achieves normalized array gains close to unity across all subcarriers, demonstrating its effectiveness in maintaining precise beam alignment over the wide frequency range. By effectively mitigating the beam split effect, AB-HBF ensures reliable performance throughout the spectrum, even as the number of antennas scales significantly. Additionally, the ability to sustain high array gains while utilizing a reduced number of RF chains highlights the framework's potential for practical, cost-efficient implementations in wideband THz massive MIMO systems.

\section{Conclusion}
This work builds upon the advancements in angular-based beamforming and hybrid precoding methodologies from mmWave systems and extends them to the unique challenges of wideband THz communication. Through rigorous theoretical analysis and simulation, we demonstrate that the proposed framework effectively reduces beam split and enhances the achievable rate, paving the way for practical, scalable THz communication solutions.
%In this paper, we propose an angular-based hybrid beamforming framework tailored for wideband THz massive MIMO systems, leveraging angular spread to address the beam split effect. Unlike previous approaches, which model angular spread as a channel characteristic, we exploit it as a design parameter to broaden beamwidth and mitigate beam misalignment. By using coarse angular information to generate subcarrier-specific beams, our method balances simplicity and performance, avoiding the need for extensive hardware modifications or complex CSI acquisition.
%This work addresses this gap by proposing an angular-based hybrid beamforming framework that leverages coarse angular information and angular spread to mitigate beam split in wideband THz massive MIMO systems. 
This approach avoids the need for complex hardware or precise channel modeling, offering a scalable and cost-effective solution for next-generation wireless networks.  
Future work will focus on optimizing the trade-off between main lobe width and interference suppression by exploring advanced precoding techniques and beam management algorithms. \vspace{-2ex}

\ifCLASSOPTIONcaptionsoff
\newpage
\fi
\bibliographystyle{IEEEtran}
\bibliography{bib_11_2023}
\balance

\end{document}